\documentclass[preprint, superscriptaddress, showpacs,preprintnumbers,amsmath,amssymb,prb]{revtex4}
\usepackage{graphicx}

\begin{document}

\thispagestyle{empty}

\title{ Casimir interaction at liquid nitrogen
temperature: Comparison between experiment and theory}

\author{R.~Castillo-Garza}
\affiliation{Department of Physics and
Astronomy, University of California, Riverside, California 92521,
USA
}
\author{J.~Xu}
\affiliation{Department of Physics and
Astronomy, University of California, Riverside, California 92521,
USA
}
\author{G.~L.~Klimchitskaya}
\affiliation{Central Astronomical Observatory
at Pulkovo of the Russian Academy of Sciences,
St.Petersburg, 196140, Russia
}
\author{V.~M.~Mostepanenko}
\affiliation{Central Astronomical Observatory
at Pulkovo of the Russian Academy of Sciences,
St.Petersburg, 196140, Russia
}
\author{U.~Mohideen}
\affiliation{Department of Physics and
Astronomy, University of California, Riverside, California 92521,
USA
}

\begin{abstract}
We have measured the normalized gradient of the Casimir force
between Au-coated surfaces of the sphere and the plate and
equivalent Casimir pressure between two parallel Au plates at
$T=77\,$K. These measurements have been performed by means of
dynamic force microscope adapted for operating at low
temperatures in the frequency shift technique.
It was shown that the measurement results at $T=77\,$K are
in a very good agreement with those at $T=300\,$K and with
computations at $T=77\,$K using both theoretical approaches to
the thermal Casimir force proposed in the literature. No thermal
effect in the Casimir pressure was observed in the limit of
experimental errors with the increase of temperature from
$T=77\,$K to $T=300\,$K. Taking this into account, we have
discussed the possible role of patch potentials
 in the
comparison between measured and calculated Casimir pressures.
\end{abstract}

\maketitle
\section{Introduction}

Rapid progress in nanotechnology has resulted in the growth of
interest in fluctuation-induced phenomena which can play a
dominant role at short separation distances below a
micrometer.\cite{1,2}
Among these phenomena a particular attention is focussed
on the Casimir effect\cite{3} which manifests itself as a force
acting between two uncharged closely spaced material boundaries.
The Casimir force is caused by zero-point and thermal
fluctuations of the electromagnetic field. In the first
approximation it depends only on the velocity of light $c$,
Planck constant $\hbar$, temperature $T$ and separation
distance $a$ between the test bodies. A more exact theory
demonstrates dependence of the Casimir force on the material
properties of the bodies \cite{4} and geometry of their
boundary surfaces.\cite{5,6}

Taking into account both fundamental interest and potential
applications in nanotechnology, many experiments have been
performed on measuring the Casimir force (see the
monograph\cite{7} and reviews\cite{8,9,10}).
Specifically, measurements with different levels of precision
were done between two metallic surfaces (see, e.g.,
Refs.~\cite{11,12,13,14,15,16}) and between a metallic and
a semiconductor surfaces (see, e.g.,
Refs.~\cite{17,18,19,20,21,22,23,24}).
Quite unexpectedly, the experimental data of many
experiments performed at room
temperature\cite{13,15,16,18,23,24,25,26}
were found to exclude the theoretical predictions taking into
account the relaxation properties of conduction electrons for
metals and the contribution of free charge carriers for
semiconductors of the dielectric type.
The same data were found to be consistent with theory
neglecting the relaxation properties of conduction electrons for
metals and the free charge carriers for
dielectric-type semiconductors
(the two experiments which claim confirmation of the role of
relaxation properties of free electrons for
metallic test bodies\cite{27,28} are critically discussed
 in the literature\cite{29,30,31,32}).
Coincident with the experimental work, it was shown\cite{33,34}
that the account of relaxation properties for metals and free
charge carriers for dielectrics leads to violation of the
Nernst heat theorem (the third law of thermodynamics) in the
Lifshitz theory. These results have led to a
critical discussion
(see, e.g., Refs.~\cite{35,36,37,38}) which continues to
the present day. Specifically, it was hypothesized\cite{39}
that the effect of large surface patches might bring the
measurement data\cite{25} in agreement with theory taking
the relaxation properties of electrons into account.
However, measurements of the Casimir force between magnetic
surfaces\cite{16,40} demonstrated that this suggestion does
not lead to the desired result.

Keeping in mind that the problems discussed above are
connected with thermal dependence of the Casimir force,
it would be elucidating to perform experiments at different
temperatures $T$.
Measurements at varying $T$ cannot entirely resolve the
theoretical problem of the Nernst theorem because experimentally
it is not possible to achieve arbitrarily low temperatures.
Such measurements, however,  can be
helpful in many aspects, specifically, for understanding
the Casimir effect between superconducting test bodies.
A  suggestion to measure the Casimir force at different
$T$ was proposed a decade
ago\cite{41} but remained unrealized due to experimental
difficulties. Recently, however, some progress in measuring
the Casimir force at low temperatures has been achieved.
Thus, the effective Casimir pressure between two parallel
plates was determined\cite{42} at $T=2.1\,$K, 4.2\,K and
77\,K by means of a micromachined oscillator.
This was done dynamically by measuring the change in
resonant frequency of an Au sphere oscillating near an Au
plate with the help of the proximity force approximation
(PFA)\cite{7,8} (note that recently the applicability of
PFA for the configuration of a sphere and a plate made of
real materials was confirmed to a high
precision\cite{43,44,45,46}).
It was shown\cite{42} that although the low temperature
data are noisier than at room temperature, the mean
measured Casimir pressures coincide at all temperatures.
Thus, it was experimentally demonstrated that within  a
wide temperature region there is no thermal effect in the
Casimir pressure exceeding the measurement errors
(no comparison with theory has been made).

In fact at short separation distances the predicted thermal
effect depends on the theoretical approach, but in all cases
it is rather small. Thus, for Au plates with neglected
relaxation properties of free electrons (the so-called
{\it plasma model approach}, see Sec.~III) spaced at
separation $a\lesssim 500\,$nm the relative thermal correction
to the Casimir pressure at $T=77\,$K is less than 0.006\%.
At $T=300\,$K the thermal correction varies from 0.02\,mPa
(0.0026\%) at $a=180\,$nm to 0.005\,mPa
(0.03\%) at $a=500\,$nm.
Such thermal corrections cannot be measured by the
presently existing experimental means.
For Au plates with included
relaxation properties of free electrons (the so-called
{\it Drude model approach}, see Sec.~III)
at $T=77\,$K the thermal correction varies from --5.8\,mPa
(--0.77\%) at $a=180\,$nm to --0.35\,mPa
(--2\%) at $a=500\,$nm.
At room temperature $T=300\,$K the thermal correction
calculated using the Drude model approach achieves
 --17\,mPa (--2.3\%) at $a=180\,$nm and --1.3\,mPa
(--7.7\%) at $a=500\,$nm.
If such corrections were really exist, they could be
observed experimentally using available setups.

Furthermore, measurements of the gradient of the Casimir
force between an Au sphere and either an Au or a doped Si plate was
performed\cite{47} at $T=4.2\,$K by means of dynamic atomic
force microscope (AFM). The experimental data were compared
with theory taking the relaxation properties of charge carriers
into account, but were found above the theoretical curve by
50\%, i.e., much more than the experimental error.
It was suggested that a possible source of disagreement with
theory is the unaccounted error in calibration of the
piezoelectric scanner extension.

Next, the dynamic AFM was adopted\cite{48} for precise
measurements of the gradient of the Casimir force between
Au surfaces of the sphere and the plate in a wide temperature
region from 5\,K to 300\,K. Unlike the previous work,\cite{47}
the developed instrument measured the extension of the
piezoelectric scanner in real time
during measurements of the Casimir
interaction. This has helped to avoid any unaccounted
systematic errors in the measurement.  Some preliminary
measurements were performed\cite{48} at $T=6.7\,$K and
demonstrated an agreement with the measurement results at
$T=300\,$K within the limits of experimental errors.

In this paper we use the instrument designed in Ref.~\cite{48},
which can operate between 5\,K and 300\,K,
for systematic measurements
of the normalized gradient of the Casimir force
between sphere and plate
at a liquid nitrogen temperature $T=77\,$K.
This is done by means of dynamic AFM operated in the frequency
shift mode.
In contrast to all previous measurements of the Casimir
interaction at low temperature, here we present the detailed
analysis of both random and systematic errors and the comparison
between the experimental data and predictions of different
theoretical approaches to the thermal Casimir force.
We demonstrate that the mean values of our data are in an
excellent agreement with the respective means at room temperature
and with predictions of both theoretical approaches at
$T=77\,$K (note that the difference between the latter at a
liquid nitrogen temperature is well below the experimental
error). The importance of these results for the problem of
thermal Casimir force is discussed.

The paper is organized as follows. In Sec.~II we briefly
consider the low-temperature setup for measurements of the
Casimir interaction and present the measurement scheme.
Section~III contains the measurement data at $T=77\,$K,
the error analysis and the comparison between experiment
and theory. In Sec.~IV the reader will find our conclusions
and discussion.

\section{Measurement setup at low temperature and
measurement scheme}

We have used dynamic AFM operated in the frequency shift
technique to measure 
the gradient of the Casimir force normalized to the sphere
radius 
between an Au-coated hollow microsphere and an Au-coated
sapphire plate at $T=77\,$K.
The experimental setup has been described in detail
elsewhere.\cite{48} In brief, however, the device we use to
measure the Casimir force gradient is a variable-temperature
force sensor (VTFS).  The apparatus is based on the AFM
 technique but particularly designed to precisely measure the
 Casimir force gradient at temperatures in the 5\,K to 300\,K
range. Here measurements are done at 77\,K. These measurements
were performed at a vacuum pressure of $\lesssim 10^{-6}\,$Torr
using an oil-free vacuum chamber.  The measurement system was
vibrationally isolated using a double-stage spring system.
Low temperatures were achieved by immersing the VTFS vacuum
chamber in a liquid nitrogen Dewar. The sensor consists of a
modified conductive Si rectangular microcantilever.\cite{48}

The Casimir force gradient was measured between a sphere and a plate
both coated with Au.  A hollow glass sphere of $R\approx 50\,\mu$m
radius made from liquid phase was attached to the end of a rectangular
Si cantilever. The sphere-cantilever system was coated uniformly with greater
than 100\,nm of Au. The low inertia of the hollow sphere leads to higher
cantilever resonant frequencies and mechanical $Q$ factors, both of which result
in improved sensitivities. Care should be taken to restrict
the Au coating to only the cantilever tip.  The cantilever with the
attached sphere had a quality factor $Q = 871$ and a resonant frequency
of $\omega_0=2\pi\times 2.766\,\mbox{kHz} =17.379\,$krad/s.

An Au-coated sapphire plate was used as the second surface. The thickness of
the Au coating on the plate was $260\pm 1\,$nm. The Au plate was mounted at a
top of a piezoelectric ceramic tube. The fabrication and characterization
methods for both surfaces have been reported in the
literature.\cite{15,48,49}
The Casimir interaction was obtained through the direct measurement of the
shift of the microcantilever's resonant frequency while the cantilever was being
lightly driven with an amplitude of 10\,nm at its resonance  frequency.\cite{50}
 The microcantilever oscillation was detected with an all-fiber
 interferometer.\cite{48} The resulting signal was analyzed and controlled
through a closed-loop feedback system provided by a phase-lock loop (PLL).
The PLL detection bandwidth was kept at 50\,Hz.  The output of the feedback loop
was the resonant-frequency shift,
$\omega_r(a)-\omega_{\rm PLL}$ where
$\omega_r(a)$ is the resonance frequency in the presence of an
external force and $\omega_{\rm PLL}$ is some frequency fixed
during the measurements. It is convenient to keep it close but
not equal to $\omega_0$.

Different voltages were applied to the plate, while the sphere remained grounded.
The resonant-frequency shifts were measured for different separations and DC
voltages between the interacting surfaces. The separation distance between the
sphere and plate was changed continuously by applying a 10\,mHz triangular voltage
signal to the piezoelectric tube holding the Au-coated sapphire plate.
The movement of the plate was  monitored in real time and calibrated with a second
all-fiber interferometer\cite{48} using $636.8\pm 0.4\,$nm wavelength light.

The relative distance moved by the piezo was calibrated with an end
cleaved fiber interferometer. We fitted the interference voltage
to third order with a cosine function using the
$\chi^2$-fitting procedure. The obtained fitting parameters
were used for calibration of piezo. First, 11 voltages from 0.56
to 0.68\,V were applied to the plate while the sphere remained
grounded. In addition, 9 repititions of 0.62\,V were also applied
to the plate, maintaining the sphere grounded.
The frequency shift was measured as a function of the distance
moved by the piezo at every 2.45\,nm for each applied voltage.
The set of measurements with 20 applications of the voltage to
the plate was repeated 3 times and one time more with 19 applied
voltages. A total of 79 runs for the frequency shift as a function
of separation were taken.
We confirmed that at large separations between $2.2\,\mu$m and
$2.6\,\mu$m the frequency shift
$\omega_r(a)-\omega_{\rm PLL}$
remains constant within the resolution limit.
This means that at separations above $2.2\,\mu$m
$\omega_r(a)=\omega_0$.

In the linear regime which occurs for small amplitudes of the
cantilever the frequency shift is given by
\begin{equation}
\Delta\omega(a)=-\frac{\omega_0}{2k}\,
\frac{\partial F_{\rm tot}(a)}{\partial a}\equiv
-\frac{\omega_0}{2k}F_{\rm tot}^{\prime}(a).
\label{eq1}
\end{equation}
\noindent
Here, $k$ is the spring constant of
the cantilever, $a$ is the absolute sphere-plate separation
distance (including the relative distance moved by the plate
piezo, $z_{\rm piezo}$, and the closest distance $z_0$),
and the frequency shift between $\omega_r(a)$ and
$\omega_0$ is determined according to the following
equation:
\begin{equation}
\Delta\omega(a)\equiv\omega_r(a)-\omega_0=
[\omega_r(a)-\omega_{\rm PLL}]-(\omega_0-\omega_{\rm PLL}).
\label{eq1a}
\end{equation}
\noindent
The total force between the sphere and the plate is the sum
of the electrostatic force $F_{\rm el}(a)$ and the Casimir force
$F_C(a)$
\begin{equation}
F_{\rm tot}(a)=F_{\rm el}(a)+F_C(a).
\label{eq2}
\end{equation}

The electric force between the sphere and the plate is expressed
as
\begin{equation}
F_{\rm el}(a)=X(a,R)(V_i-V_0)^2,
\label{eq3}
\end{equation}
\noindent
where $V_0$ is the residual potential difference between the
bodies,
which can be caused by the various connections and by different work
functions of a sphere and a plate materials, and $X(a,R)$ is an
explicitly known function.\cite{7,8}
Using the expansion of this function in powers of a small
parameter $a/R$ obtained in the literature,\cite{7,8,17}
we can rewrite Eq.~(\ref{eq1}) in the form
\begin{eqnarray}
&&
\Delta\omega(a)=-\frac{\omega_0}{2k}F_C^{\prime}(a)
-\frac{\omega_0}{2k}\,\frac{\partial X(a,R)}{\partial a}
(V_i-V_0)^2
\label{eq4}\\
&&~~
=
-\frac{\tilde{C}}{2\pi R}F_C^{\prime}(a)
-\frac{\tilde{C}\epsilon_0}{2a^2}\left[1-
\sum_{i=1}^{6}ic_i\left(\frac{a}{R}\right)^{i+1}\right]
(V_i-V_0)^2.
\nonumber
\end{eqnarray}
\noindent
Here, $\tilde{C}\equiv\pi\omega_0 R/k$ is the calibration constant
 [we label $C$ with a tilde to indicate the difference with
$C=\omega_0/(2k)$ used in the literature\cite{15,16,40}],
$\epsilon_0$ is the permittivity of free space and the numerical
coefficients $c_i$ can be found in Refs.~\cite{7,8,17}.

Any mechanical or thermal drift of the piezo leading to a change
in the sphere-plate separation distance was found to be
$1.11226 \pm 0.00455\,$nm/hour.  The corresponding corrections
to this small drift were done as reported previously.\cite{15}
After applying the drift correction, the residual potential
$V_0$ was found at each separation from the parabolic dependence
of the measured frequency shift on $V_i$. The value of $V_0$
at a fixed separation can be identified as the position of the
parabola maximum. The obtained $V_0$ as a function of separation
is shown by dots in Fig.~\ref{fg1}. As is seen in the figure,
the mean $V_0=0.6125\pm 0.0015\,$V is independent of separation
over the entire measurement range. To quantify this observation,
we have performed the best fit of $V_0$ to the straight line
leaving its slope as a free parameter (solid line in
Fig.~\ref{fg1}).
It was found that the slope is equal to
$-5.7\times 10^{-7}\,$V/nm, i.e., the independence of $V_0$
on separation was confirmed to a high precision.

The curvature of the parabolas of the measured frequency shift as
a function of $V_i$ corresponds to the spatial dependence of the
electrostatic force and the force calibration constant
$\tilde{C}$.
In accordance with Eq.~(\ref{eq4}), this parabola curvature was
fitted to the quantity
\begin{equation}
\beta\equiv\frac{\tilde{C}\epsilon_0}{2a^2}\left[1-
\sum_{i=1}^{6}ic_i\left(\frac{a}{R}\right)^{i+1}\right]
\label{eq5}
\end{equation}
\noindent
in order to determine the calibration constant $\tilde{C}$ and
the closest mean sphere-plate separation $z_0$ taking into
account that $a=z_0+z_{\rm piezo}$. The fitting procedure was
repeated by keeping the start point fixed at the closest
separation, while the end point $z_{\rm end}$ measured from the
closest separation was varied over a wide range.
Similar to Ref.~\cite{15}, $z_0$ and $\tilde{C}$ so determined
were shown to be independent on  $z_{\rm end}$.
The obtained values of the calibration parameters are
$z_0=184.4\pm 1.5\,$nm and
$\tilde{C}=78.2\pm 1.0\,\mbox{rad\,m}^2/${N\,s).
The errors are determined by the systematic errors in the fit.

The Casimir force and electrostatic force between sphere and plate
are both attractive and cause the cantilever to bend towards the
plate by around 1\,nm at the maximum applied voltage
at the closest separation.
 Here, there was no real time correction of the cantilever bending
 using a proportional integral derivative loop,
as reported in Ref.~\cite{15}.
 Thus, for a precise measurement, all the values of absolute
separations $a$ have to be
corrected for this bending.
The correction was done in the following manner.
By integrating Eq.~(\ref{eq1}) from $a$ to $\infty$
we can get
\begin{equation}
F_{\rm tot}(a)=\frac{2k}{\omega_0}\int_{a}^{\infty}\!\!
\Delta\omega(z)\,dz,
\label{eq6}
\end{equation}
\noindent
where $a<L=2.2\,\mu$m and $L$ is the largest distance at which
the frequency shift was measured.
Keeping in mind that at $a\geq L$ the Casimir force is nearly
equal to zero, from Eq.~(\ref{eq2}) we obtain
$F_{\rm tot}(a)\approx F_{\rm el}(a)$ and then
\begin{equation}
\frac{2k}{\omega_0}\int_{L}^{\infty}\!\!
\Delta\omega(z)\,dz\approx
F_{\rm el}(L).
\label{eq7}
\end{equation}
\noindent
Subdividing the integration region in Eq.~(\ref{eq6}) in two
subregions $[a,L]$ and $[L,\infty)$ and using Eq.~(\ref{eq7}),
we rewrite Eq.~(\ref{eq6}) in the form
\begin{equation}
F_{\rm tot}(a)\approx\frac{2k}{\omega_0}\int_{a}^{L}\!\!
\Delta\omega(z)\,dz+F_{\rm el}(L).
\label{eq8}
\end{equation}
\noindent
Finally, with the help of Eq.~(\ref{eq3}), Hooke's law
and the expansion of the function $X(a,R)$ in powers of a small
parameter,\cite{7,8,17} one arrives to the following bending
distance of the cantilever:
\begin{eqnarray}
&&
b(a)=\frac{F_{\rm tot}(a)}{k}=\frac{2}{\omega_0}\int_{a}^{L}\!\!
\Delta\omega(z)\,dz+\frac{F_{\rm el}(L)}{k}
\label{eq9} \\
&&~~~
=\frac{2}{\omega_0}\int_{a}^{L}\!\!
\Delta\omega(z)\,dz-
\frac{\tilde{C}\epsilon_0}{\omega_0 L}
\left[1+
\sum_{i=0}^{6}c_i\left(\frac{L}{R}\right)^{i+1}\right]
(V_i-V_0)^2.
\nonumber
\end{eqnarray}

Now we can use Eq.~(\ref{eq9}) to correct all the values
of absolute separation $a=a_0=z_0+z_{\rm piezo}$
at any applied voltage
for a bending of the cantilever by means of an iteration
procedure. First we substitute $a_0$ (zero iteration) in
Eq.~(\ref{eq9}) and calculate $b(a_0)$ at any $z_{\rm piezo}$.
The first iteration of absolute separations is defined as
$a_1=a_0-b(a_0)$. Substituting this in Eq.~(\ref{eq9}), we
find $b(a_1)$ etc.
In the iteration number $i$ we have $a_i=a_{i-1}-b(a_{i-1})$.
This is iterated until the obtained value converge
at any $z_{\rm piezo}$ and any applied voltage.
 Note that for all separations and applied voltages
the correction due to the
bending of the cantilever is smaller than 1\,nm and
decreases with increasing separation. Thus, it is
below the error in the determination
of absolute separations. The obtained values of all calibration
parameters $V_0$, $\tilde{C}$, $z_0$ and absolute separations
$a$ can now be used to perform an independent measurement
of the Casimir interaction at the liquid nitrogen temperature.

\section{Measurement data, error analysis and comparison with
theory}

As discussed in Sec.~II, the frequency shift $\Delta\omega$
caused by the combined action of the electric and Casimir forces
was measured as a function of separation 79 times with
different applied voltages over the separation region from
187\,nm to $2.2\,\mu$m.
Then the normalized gradients of the Casimir force,
$F_C^{\prime}(a)/(2\pi R)$, at different separations were
found from Eq.~(\ref{eq4}). For comparison purposes, it is
convenient to recalculate the normalized gradients of the
Casimir force in sphere-plate geometry into the Casimir
pressure $P_C(a)$ between two Au-coated plane parallel
plates. This can be done by means of the PFA\cite{7,8}
\begin{equation}
F_C(a)=2\pi R E_C(a),
\label{eq10}
\end{equation}
\noindent
where $E_C(a)$ is the Casimir energy per unit area of two
parallel plates. The negative differentiation of both sides
of Eq.~(\ref{eq10}) leads to the desired result
\begin{equation}
P_C(a)=-\frac{1}{2\pi R} F_C^{\prime}(a),
\label{eq11}
\end{equation}
\noindent
i.e., the Casimir pressure coincides with the negative
normalized gradient. At separations of several hundred
nanometers, important for this experiment, the error
introduced by the use of PFA does not exceed a fraction of
a percent\cite{43,44,45,46}, i.e., much less than the
experimental error (see below).

In Fig.~\ref{fg2}
all 79 individual values of the Casimir pressure measured
at each point are shown
 at short separation distances as gray dots with a step of
2.45\,nm.
In the same figure the mean values of the measured Casimir
pressure at each separation are interpolated and presented
as the solid line.

The statistical properties of the measurement data for the
Casimir pressure are characterized by the histogram in
Fig.~\ref{fg3} plotted at $a=187\,$nm. Here, $f$ is the
fraction of 79 data points having the force values in the bin
indicated by the respective vertical lines. The histogram is
described by the Gaussian distribution with the standard
deviation equal to $\sigma_{P_C}=11.5\,$mPa and the mean
Casimir pressure $\bar{P}_C=-650.65\,$mPa
(see also below for the comparison with predictions of
different theoretical approaches indicated in Fig.~\ref{fg3}
by the solid and dashed vertical lines).

It is instructive to compare the measurement data at $T=77\,$K
and $T=300\,$K. This is done in Fig.~\ref{fg4}(a,b) where the
two histograms are presented at $a=234\,$nm, $T=77\,$K
(this work) and at $a=235\,$nm, $T=300\,$K (by the results of
Ref.~\cite{15}), respectively. The respective standard
deviations and mean Casimir pressures are
$\sigma_{P_C}=10.5\,$mPa,  $\bar{P}_C^{\rm expt}=-287.87\,$mPa
[$T=77\,$K, Fig.~\ref{fg4}(a)] and
$\sigma_{P_C}=2.75\,$mPa,  $\bar{P}_C^{\rm expt}=-284.17\,$mPa
[$T=300\,$K, Fig.~\ref{fg4}(b)].
As can be seen from the comparison of Fig.~\ref{fg4}(a) and
Fig.~\ref{fg4}(b), at $a\approx 235\,$nm the mean Casimir
pressures measured at the liquid nitrogen and room temperatures are
in very good agreement although the data at $T=77\,$K are less
precise.

To compare our experimental data with the data of other
experiments and with theory over wide separation regions,
we first analyze the experimental errors. The random error in the
measured Casimir pressure is calculated by using Student's
distribution. As a function of separation, the random error
determined at a 67\% confidence level is shown by the dotted line
in Fig.~\ref{fg5}. The largest value of the random error equal to
1.3\,mPa is achieved at the shortest separation  $a=187\,$nm.
Then it decreases down to 0.8\,mPa when separation increases up to
600\,nm and preserves this value at larger separations.
The systematic error in this experiment is determined by the
instrumental noise including the background noise level and
by the errors in calibration. In Fig.~\ref{fg5} the systematic
error is shown by the long-dashed line.
As can be seen in this figure, the systematic error achieves
the largest value of 6.8\,mPa at $a=187\,$nm, decreases to 4.0\,mPa
at $a=250\,$nm and is equal to 3.4\,mPa at all separation
distances larger than 340\,nm.
We emphasize that the systematic error in this experiment is
larger than
in previously performed experiment\cite{15} by means of AFM
at $T=300\,$K (where it was approximately equal to
1.8--1.9\,mPa).
The significant increase of the systematic error in the
low-temperature setup, as compared to the room temperature,
is due to internal vibrations when the cryogenic liquid is
present.\cite{42}
By adding the random and systematic errors in quadrature,
we obtain the total experimental error in the measured
Casimir pressure determined at a 67\% confidence level.
It is shown by the solid line in Fig.~\ref{fg5}.
The total error is mostly determined by the systematic error.
It is equal to 6.9\,mPa at $a=187\,$nm, decreases to 4.1\,mPa
at $a=250\,$nm, and preserves the value of 3.5\,mPa
at all separations exceeding 450\,nm.

Now we plot our mean experimental data as crosses in
Fig.~\ref{fg6}(a,b), where the vertical arms are equal to the
total experimental error in the measured Casimir pressure and
the horizontal arms are equal to the error in the measurement of
separations $\Delta a=\Delta z_0=1.5\,$nm
(the error in $z_{\rm piezo}$ is negligibly small).
The separation regions are chosen for the  comparison with the
mean measured Casimir pressures of Ref.~\cite{15} and
Refs.~\cite{25,26} shown by the solid lines in
Figs.~\ref{fg6}(a) and \ref{fg6}(b), respectively.
As can be seen in Fig.~\ref{fg6}(a,b), our measurement data
at $T=77K$ are in a very good agreement in the limits of
experimental errors with the measurement data of Ref.~\cite{15}
obtained by means of AFM and Refs.~\cite{25,26} obtained by
means of micromachined oscillator at $T=300\,$K.
This demonstrates that there is a mutual agreement between all
these experiments  and that there is no thermal effect exceeding
the measurement errors when the temperature increases from
77\,K to 300\,K.

Next we compare the measurement data for the mean Casimir
pressure with theory. The thickness of Au coatings is large
enough to consider our effective parallel plates as
semispaces made of gold.\cite{7}
The Lifshitz formula for the Casimir pressure between two
semispaces at temperature $T$ is given by\cite{4,7,8,9}
\begin{equation}
P_C(a)=-\frac{k_BT}{\pi}
\sum_{l=0}^{\infty}{\vphantom{\sum}}^{\prime}
\int_{0}^{\infty}\! q_lk_{\bot}dk_{\bot}
\sum_{\alpha}\frac{r_{\alpha}^2}{e^{2aq_l}-r_{\alpha}^2},
\label{eq12}
\end{equation}
\noindent
where $k_B$ is the Boltzmann constant and the laboratory
temperature is $T=77\,$K. The quantity
$q_l^2=k_{\bot}^2+\xi_l^2/c^2$ where $k_{\bot}$ is the
projection of the wave vector on the plane of plates,
and $\xi_l=2\pi k_BTl/\hbar$ with $l=0,\,1,\,2,\,\ldots$
are the Matsubara frequencies. The prime near the first
summation sign multiplies the term with $l=0$ by 1/2,
and the second summation sign summarizes over the
transverse magnetic ($\alpha={\rm TM}$) and transverse
electric ($\alpha={\rm TE}$) polarizations of the
electromagnetic field. The reflection coefficients
$r_{\alpha}$ are calculated along the imaginary frequency
azis. They are given by
\begin{eqnarray}
&&
r_{\rm TM}\equiv r_{\rm TM}(i\xi_l,k_{\bot})=
\frac{\varepsilon(i\xi_l)q_l-k_l}{\varepsilon(i\xi_l)q_l+k_l},
\nonumber \\
&&
r_{\rm TE}\equiv r_{\rm TE}(i\xi_l,k_{\bot})=
\frac{q_l-k_l}{q_l+k_l},
\label{eq13}
\end{eqnarray}
\noindent
where the dielectric permittivity of Au calculated at the
imaginary Matsubara frequencies is $\varepsilon(i\xi_l)$ and
\begin{equation}
k_l=\left[k_{\bot}^2+\varepsilon(i\xi_l)
\frac{\xi_l^2}{c^2}\right]^{1/2}.
\label{eq14}
\end{equation}

As discussed in Sec.~I, there are two approaches on how to apply
the Lifshitz theory to metallic bodies. The Drude model
approach\cite{7,8,35,36} takes into account the relaxation
properties of conduction electrons. In the framework of this
approach, the optical data of boundary metal are extrapolated
down to zero frequency by means of the Drude model and are
used to calculate $\varepsilon(i\xi_l)$ with the help of the
dispersion relation. The tabulated optical data\cite{51}
of Au are well extrapolated\cite{52} by the Drude model with
the plasma frequency $\omega_p=9.0\,$eV and relaxation
parameter $\gamma=0.035\,$eV determined at $T=300\,$K.
[Note that although $\omega_p$ is temperature-independent,
the relaxation parameter decreases with decreasing
temperature, so that $\gamma(77\,\mbox{K})\approx0.9\,$meV.]
The consistency of this extrapolation was recently confirmed
by using the weighted Kramers-Kronig relations.\cite{53}
The immediately measured optical data for Au films, similar to
those used in experiments of Refs.~\cite{15,25,26}, by means
of ellipsometry lead to the same Casimir pressures as the
tabulated optical data.\cite{42}
In the framework of the plasma model approach, the same
optical data with the contribution of free charge carriers
subtracted are extrapolated down to zero frequency by means
of the plasma model with the same $\omega_p$.

We have calculated the Casimir pressure at $T=77\,$K using
Eqs.~(\ref{eq12})--(\ref{eq14}) over the entire measurement region
from 187\,nm to $2\,\mu$m in the framework of both approaches
(note that the surface roughness in this experiment contributes
a small fraction of a percent and can be neglected in
comparison to the error bars\cite{15}).
The computational results are presented by the solid lines
in Fig.~\ref{fg7}(a) within the separation region from 187 to
300\,nm and in Fig.~\ref{fg7}(b) within the region from 300 to
500\,nm. We emphasize that with the scale used the differences
between predictions of the Drude and plasma model approaches
are below resolution. In the same figures, the experimental
data are shown as crosses, as discussed above.
From Fig.~\ref{fg7}(a,b) it can be seen that the measured mean
Casimir pressures are in good agreement with theory.
In order to trace the differences between the predictions of the
Drude and plasma model approaches, we return to Figs.~\ref{fg3}
and \ref{fg4} where both predictions are shown by the dashed and
solid vertical lines, respectively. In Fig.~\ref{fg3}
($a=187\,$nm),
$P_D^{\rm th}=-652.25\,$mPa and
$P_p^{\rm th}=-653.61\,$mPa
leading to only a 1.36\,mPa difference. This is much smaller than
the total experimental error at this separation (6.9\,mPa) and
also smaller than the theoretical error equal to 3.3\,mPa.

To compare the predictions of both computational approaches
at $T=77\,$K and $T=300\,$K one should look to Figs.~\ref{fg4}(a)
and \ref{fg4}(b), respectively.
In  Fig.~\ref{fg4}(a), one has
$P_D^{\rm th}(T=77\,\mbox{K})=-286.53\,$mPa and
$P_p^{\rm th}(T=77\,\mbox{K})=-288.16\,$mPa.
This leads to a 1.63\,mPa difference still smaller than the
total experimental error
$\Delta^{\!\rm tot}\bar{P}_C^{\rm expt}(T=77\,\mbox{K})=4.4\,$mPa.
In contrast in Fig.~\ref{fg4}(b)
plotted at $T=300\,$K,
$P_D^{\rm th}(T=300\,\mbox{K})=-273.99\,$mPa and
$P_p^{\rm th}(T=300\,\mbox{K})=-283.35\,$mPa
leading to the difference of 9.36\,mPa much larger than the total
experimental error in this experiment equal to
$\Delta^{\!\rm tot}\bar{P}_C^{\rm expt}(T=300\,\mbox{K})=1.9\,$mPa.
[Note that the difference between the plasma model predictions
in Figs.~\ref{fg4}(a) and \ref{fg4}(b) is equal to --4.81\,mPa;
the major part of this --4.82\,mPa, is due to the change of
separation from 234 to 235\,nm and only 0.01\,mPa is due
to the change of temperature from 77\,K to 300\,K.
This reflects the fact that in the framework of the
plasma model approach the thermal effect at short separations
is very small.]
We emphasize also that
$P_D^{\rm th}(T=300\,\mbox{K})-P_p^{\rm th}(T=300\,\mbox{K})
>\Delta^{\!\rm tot}P_C^{\rm expt}(T=77\,\mbox{K})$.
As can be seen in Fig.~\ref{fg4}(b), the prediction of the
plasma model approach is in a very good agreement with the
measurement data, whereas the prediction of the Drude model
approach is excluded by the data.

To get an idea on the comparison between experiment and theory at
larger separation distances, in Fig.~\ref{fg8} we show the
theoretical predictions for the Casimir pressure at $T=77\,$K
within the region from 500 to 2000\,nm by the white bands
(for the scale used the difference between the predictions of both
approaches is again below the resolution).
The results of all individual pressure measurements are indicated
as dots in Fig.~\ref{fg8}(a), whereas mean Casimir pressures with
their total experimental errors are shown as crosses in
Fig.~\ref{fg8}(b). As can be seen in Fig.~\ref{fg8}(a,b), the data
are meaningful up to approximately 700\,nm and at $a>1\,\mu$m
the measured signal is averaged to zero.
Thus, there is no offset in the calibration of our setup.

\section{Conclusions and discussion}

In the foregoing we have presented the measurement results for
the normalized gradient of the Casimir force at liquid
nitrogen temperature between Au-coated surfaces of a sphere and
a plate. Using the PFA, these results were recalculated into the
Casimir pressure between two Au-coated plates and compared with
theoretical predictions of the Drude and plasma model approaches
at $T=77\,$K and with the Casimir pressure measured at $T=300\,$K.
It was found that although measurements at cryogenic
temperatures are burdened with a larger systematic error, they
are in a very good agreement with the measurement results
at $T=300\,$K and with theoretical predictions of both approaches
at $T=77\,$K. We have calculated the differences between the
predictions  of the Drude and plasma model approaches
at $T=77\,$K and shown that they are below the instrumental
sensitivity. We have also traced that with the increase of
temperature up to $T=300\,$K the difference between the
predictions of both approaches exceeds the
increased instrumental sensitivity, so that the measurement
data exclude the Drude model approach and are consistent with
the plasma model approach.

The performed cryogenic measurements have been possible due to
the use of dymanic AFM which can operate in high vacuum
environments and temperatures between 5\,K and 300\,K.
This setup opens prospective opportunities for measuring the
Casimir interaction between different samples at variable
temperature and for comparison of the obtained measurement
data with theory. As is shown in this paper, there are no
detectable changes in the thermal Casimir pressure when the
measurement data taken at $T=77\,$K and $T=300\,$K are
compared.

The obtained results shed some additional light on the possible role of
surface patches in the comparison between measured and
calculated Casimir pressures.\cite{39}
As mentioned in Sec.~I, measurements of the Casimir
interaction between magnetic surfaces\cite{16,40}
are not compatible with this hypothesis.
Under some assumptions a similar conclusion that there is no
significant contribution of patches can be arrived for
nonmagnetic (Au) surfaces when the measurement data at
two different temperatures are compared.
To see this, let us assume for a while that at $T=300\,$K
there is large contribution of patch potentials to the measured
pressure equal to
$|P_{\rm patch}|=|P_p^{\rm th }(T=300\,\mbox{K})-
P_D^{\rm th}(T=300\,\mbox{K})|\approx 9.36\,\mbox{mPa}
>\Delta^{\!\rm tot}\bar{P}_C^{\rm expt}(T=300\,\mbox{K})
=1.9\,$mPa
at $a=235\,$nm (see Sec.~III), such that
${P}_D^{\rm th}(T=300\,\mbox{K})+P_{\rm patch}
\approx \bar{P}^{\rm expt}(T=300\,\mbox{K})$.
This, however, comes into conflict with the fact that at $T=77\,$K
our measurement data are in a good agreement with
theoretical predictions of
both the Drude and the plasma model approaches, which are very
close:
$P_D^{\rm th}(T=77\,\mbox{K})-P_p^{\rm th}(T=77\,\mbox{K})
\approx 1.63\,\mbox{mPa}
<\Delta^{\!\rm tot}\bar{P}_C^{\rm expt}(T=77\,\mbox{K})
=4.4\,$mPa (see Sec.~III).
Really, if one would add the patch effect of --9.36\,mPa
to $P_D^{\rm th}(T=77\,\mbox{K})$, the obtained result of
--295.9\,mPa differs from
$\bar{P}_C^{\rm expt}(T=77\,\mbox{K})=-287.9\,$mPa
by 8\,mPa, i.e., larger than the total experimental error
determined at a 67\% confidence level.
Thus, under the assumption that the patch effect does not
depend on temperature (i.e., it is the same at $T=300\,$K and
$T=77\,$K) the presence of a large patch effect brings the
experimental data at $T=77\,$K in disagreement with both
the Drude and plasma model approaches to the Casimir force
(the possibility of temperature-dependent patch potentials
awaits further investigation).

To conclude, we express a hope that the fundamental understanding
of the thermal Casimir interaction between real material bodies
will be achieved in the near future.

\section*{Acknowledgments}

This work was supported by the DOE grant DEF010204ER46131 (U.M.).

\begin{figure}[b]
\vspace*{-3cm}
\centerline{\hspace*{3cm}
\includegraphics{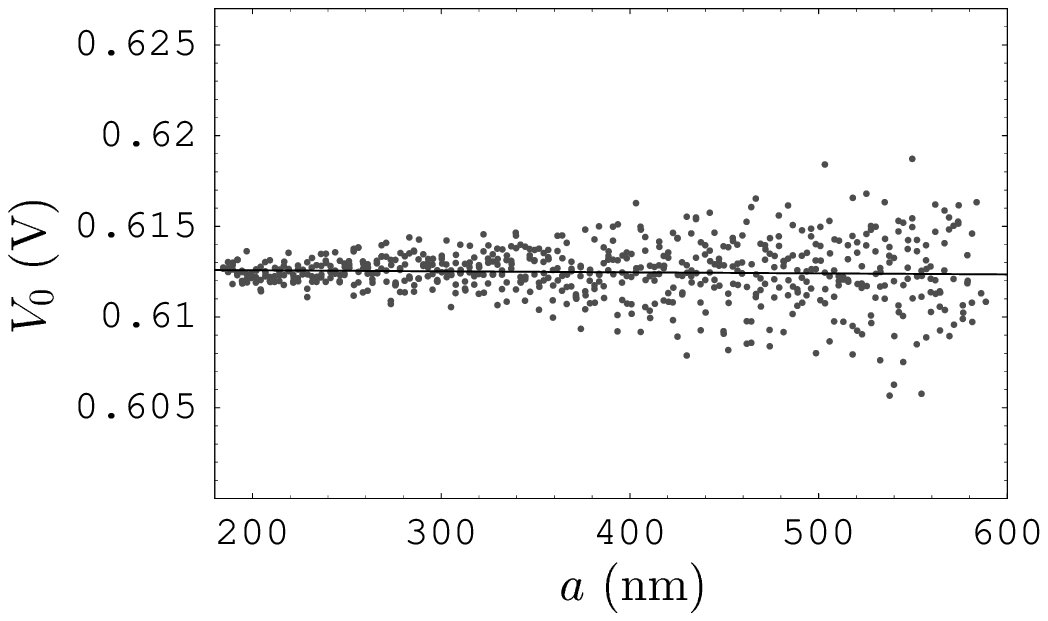}
}
\vspace*{-14cm}
\caption{\label{fg1}
The residual potential difference $V_0$ between an Au-coated
sphere and  an Au-coated plate at $T=77\,$K
as a function of separation.
The best fit of $V_0$ to the straight line is shown by the
solid line.
}
\end{figure}
\begin{figure}[b]
\vspace*{-3cm}
\centerline{\hspace*{3cm}
\includegraphics{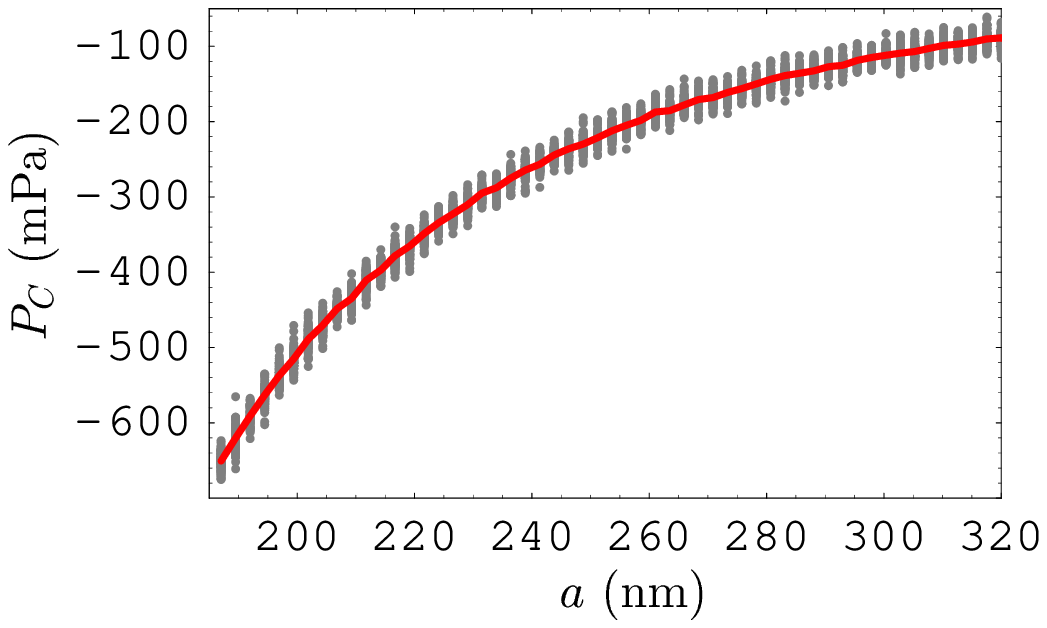}
}
\vspace*{-14cm}
\caption{\label{fg2}(Color online)
All 79 individual values of the Casimir pressure
at $T=77\,$K  measured
at each point are shown as gray dots with a step of
2.45\,nm. The mean measured Casimir pressure
as a function of separation is indicated by the solid line.
}
\end{figure}
\begin{figure}[b]
\vspace*{-1cm}
\centerline{\hspace*{3cm}
\includegraphics{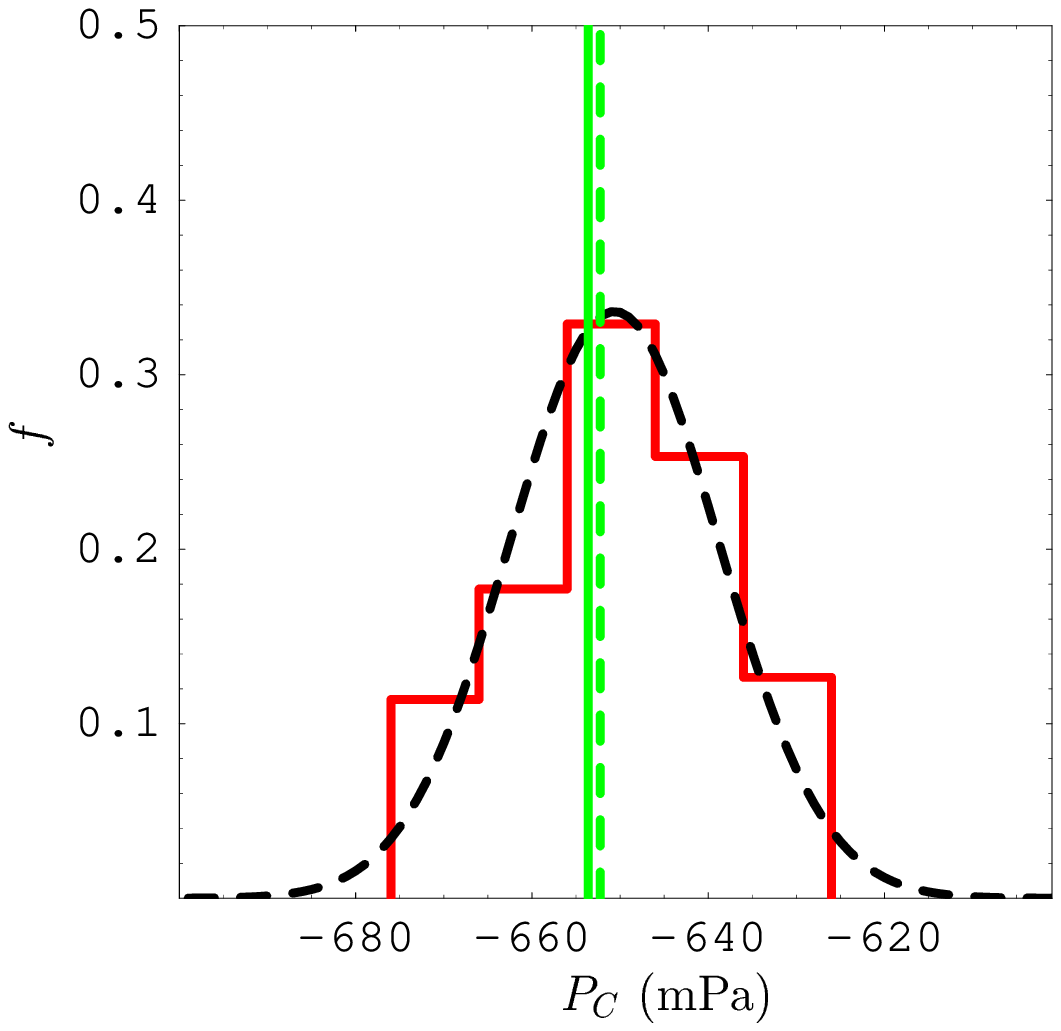}
}
\vspace*{-16cm}
\caption{\label{fg3}(Color online)
The histogram for the measured Casimir pressure
at $T=77\,$K  at the
separation $a=187\,$nm. The corresponding Gaussian
distribution is shown by the dashed curve.
The solid and dashed vertical lines indicate the
theoretical predictions from the plasma and Drude model
approaches, respectively (see text for further discussion).
}
\end{figure}
\begin{figure}[b]
\vspace*{1cm}
\centerline{\hspace*{3cm}
\includegraphics{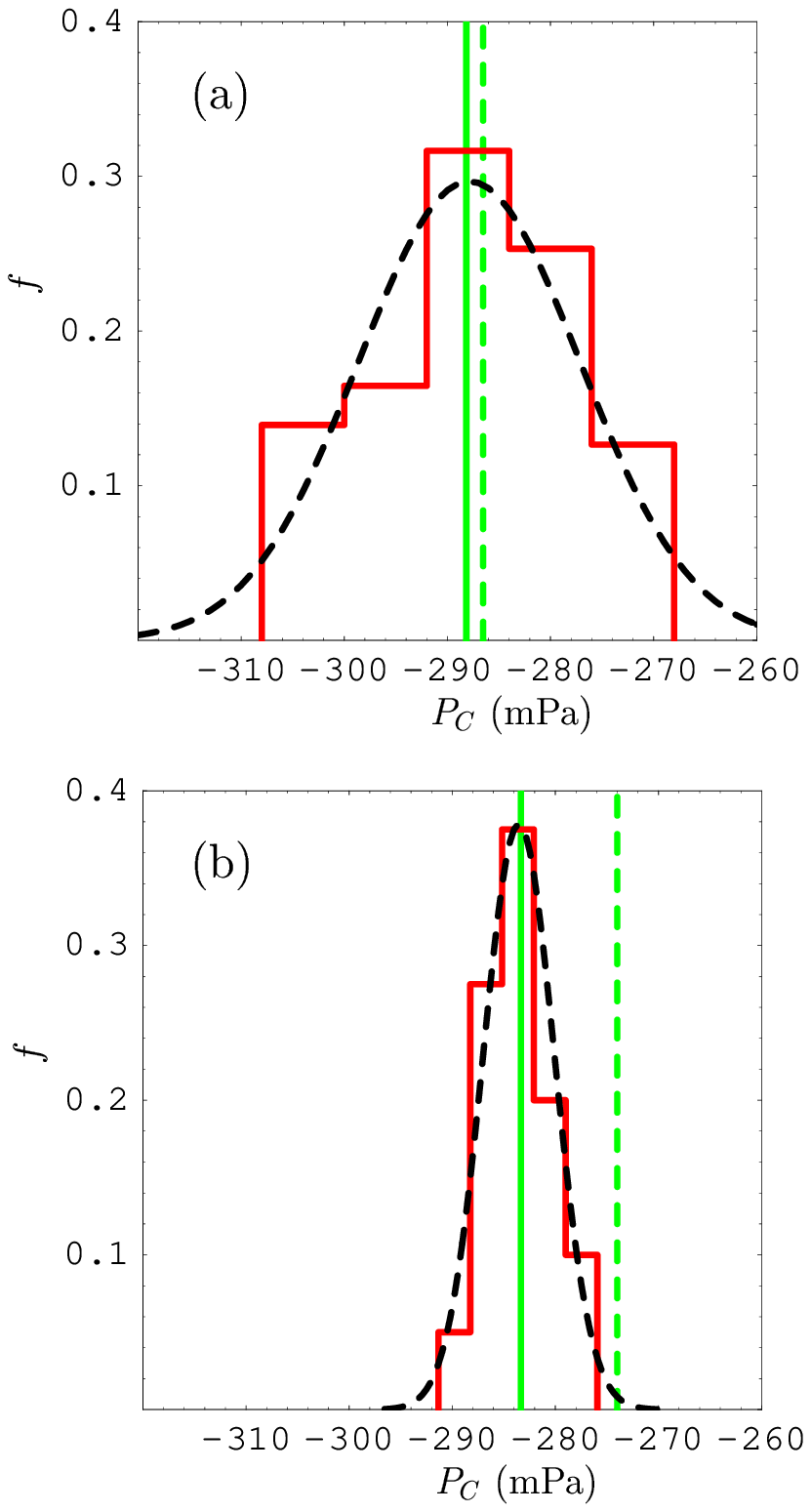}
}
\vspace*{-13cm}
\caption{\label{fg4}(Color online)
The histogram for the measured Casimir pressure
(a) at $a=234\,$nm, $T=77\,$K and
(b) $a=235\,$nm, $T=300\,$K
measured in this work and in Ref.~\cite{15},
respectively.
The corresponding Gaussian
distributions are shown by the dashed curves.
The solid and dashed vertical lines indicate the
theoretical predictions from the plasma and Drude model
approaches, respectively (see text for further discussion).
}
\end{figure}
\begin{figure}[b]
\vspace*{-3cm}
\centerline{\hspace*{3cm}
\includegraphics{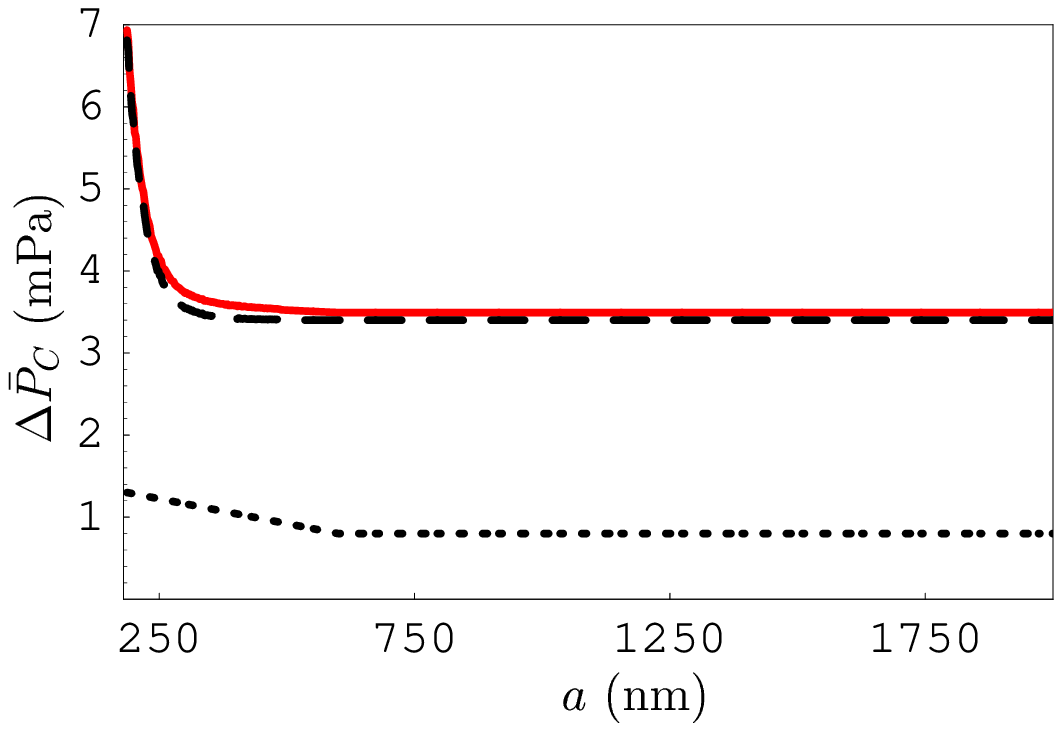}
}
\vspace*{-14cm}
\caption{\label{fg5}(Color online)
The random, systematic and total experimental errors
in the measured mean Casimir pressure
at 77\,K determined at a 67\%
confidence level are shown by the dotted, dashed, and
solid lines, respectively.
}
\end{figure}
\begin{figure}[b]
\vspace*{1cm}
\centerline{\hspace*{3cm}
\includegraphics{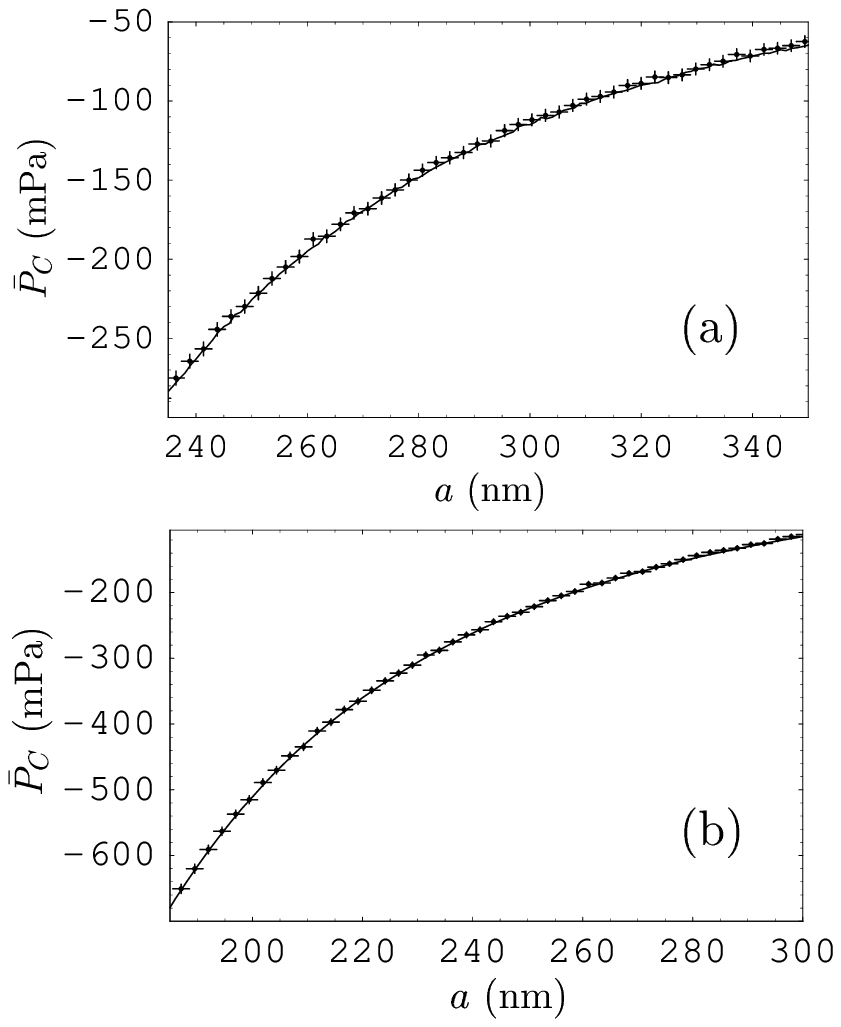}
}
\vspace*{-17cm}
\caption{\label{fg6}
The mean Casimir pressures previously measured at $T=300\,$K
(a) by means of dynamic AFM\cite{15} and
(b) by means of micromachined oscillator\cite{25,26} are shown
by the solid lines as functions of separation.
The mean Casimir pressures measured in this work at $T=77\,$K
with their total experimental errors determined at a 67\%
confidence level are indicated as crosses.
}
\end{figure}
\begin{figure}[b]
\vspace*{1cm}
\centerline{\hspace*{3cm}
\includegraphics{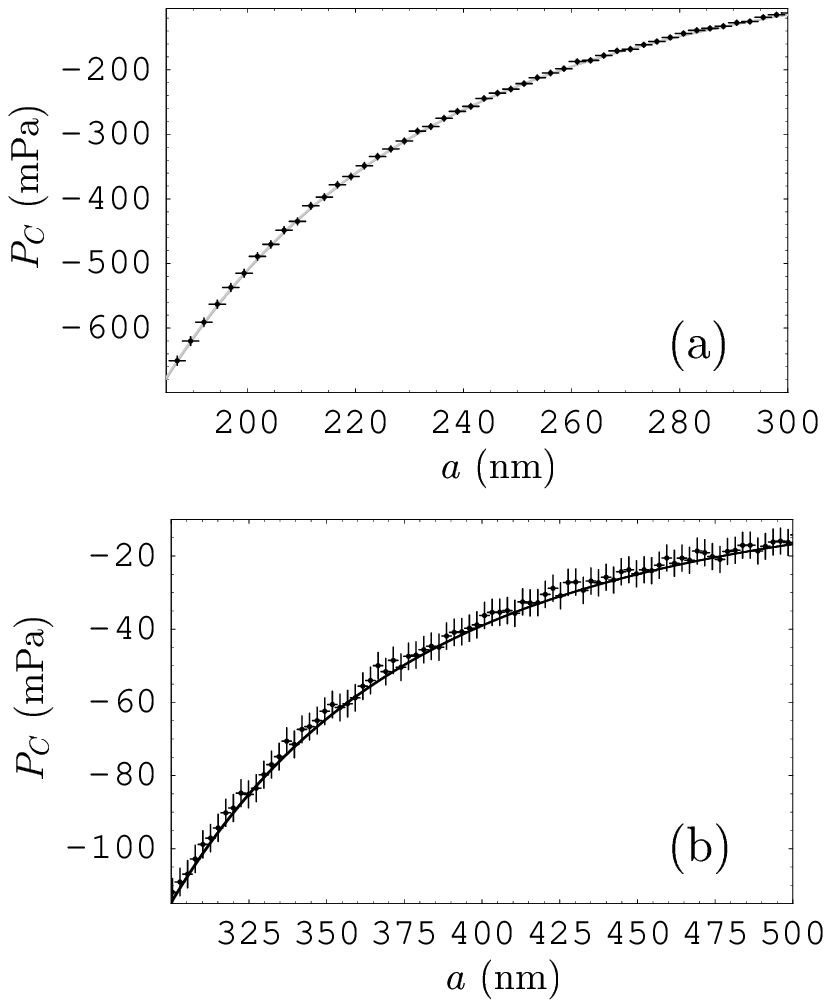}
}
\vspace*{-17cm}
\caption{\label{fg7}
Comparison between the mean experimental data for the
Casimir pressure at $T=77\,$K (crosses plotted at a 67\%
confidence level) and the common theoretical prediction
of the Drude and plasma model approaches at $T=77\,$K
(solid lines) within the separation region
(a) from 187 to 300\,nm and
(b) from 300 to 500\,nm.
}
\end{figure}
\begin{figure}[b]
\vspace*{1cm}
\centerline{\hspace*{3cm}
\includegraphics{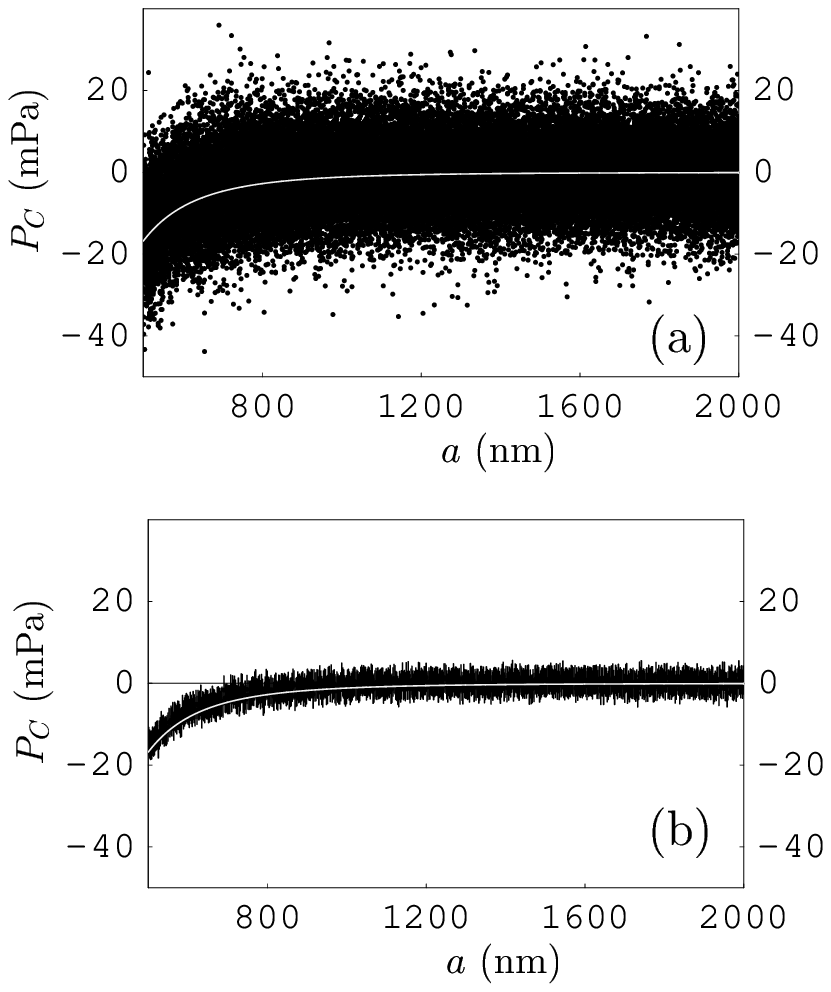}
}
\vspace*{-17cm}
\caption{\label{fg8}
Comparison between (a) the individual and (b) mean
experimental data for the Casimir pressure at $T=77\,$K
(dots and crosses, respectively)
and the common theoretical prediction
of the Drude and plasma model approaches at $T=77\,$K
(white lines).
}
\end{figure}
\end{document}